\documentclass[prd,twocolumn,floatfix,preprintnumbers,letterpaper]{revtex4}
\usepackage{amsmath,amssymb,graphicx}

\begin{document}

\title{Observational constraint on the varying speed of light theory}

\author{Jing-Zhao Qi}
\affiliation{Department of Physics, Institute of Theoretical Physics, Beijing Normal University, Beijing, 100875, China}
\author{Ming-Jian Zhang}
\author{Wen-Biao Liu}
\email{wbliu@bnu.edu.cn}
\affiliation{Department of Physics, Institute of Theoretical Physics, Beijing Normal University, Beijing, 100875, China}

\begin{abstract}
The varying speed of light theory is controversial. It succeeds in explaining some cosmological problems, but on the other hand it is excluded by mainstream physics because it will shake the foundation of physics. In the present paper, we devote ourselves to testing whether the speed of light is varying from the observational data of the type Ia supernova, baryon acoustic oscillation, observational $H(z)$ data and cosmic microwave background. We select the common form $c(t)=c_0a^n(t)$ with the contribution of dark energy and matter, where $c_0$ is the current value of speed of light and $n$ is a constant, and consequently we construct a varying speed of light dark energy model (VSLDE). The combined observational data show a trivial constraint $n=-0.0033 \pm 0.0045$ at 68.3\% confidence level, which indicates that the speed of light may be a constant with high significance. By reconstructing the time variable $c(t)$, we find that the speed of
light almost has no variation for redshift $z < 10^{-1}$. For high-$z$ observations, they are more sensitive to the VSLDE model, but the variation of speed of light is only in order of $10^{-2}$. We also introduce the geometrical diagnostic $Om (z)$ to show the difference between the VSLDE and $\Lambda$ cold dark matter ($\Lambda$CDM) models. The result shows that the current data make it difficult to differentiate them. All the results show that the observational data favor the constant speed of light.

\end{abstract}

\maketitle

\section{Introduction}  \label{introduction}

The ``standard big bang" model of the universe has been recognized by most of the cosmologists. However, there still exist some puzzles to be  clarified. The inflation model as a paradigm to solve the cosmological puzzles has made considerable success in understanding the Universe, especially the recent detection of the B-mode from cosmic microwave background (CMB) by the BICEP2 group \cite{ade2014bicep2} strongly consolidates this model. However, the primordial seeds and some other structures should be built in the initial conditions. Different from the inflation model, \citet{albrecht1999time} introduced a model of time-variable speed of light $c$ in the early universe to solve the cosmological horizon, flatness, and cosmological constant problems. This type of model is usually called the varying speed of light (VSL) theory and gradually becomes an alternative to the inflation. In the VSL theory, the radiation also dominates the Universe at early times. Einstein's gravity is also valid. Therefore, the cosmic geometry and expansion factor are the same as in the standard big bang model. However, free falling observers would measure that the local speed of light varies with cosmic time. Since then, this theory was further studied by Moffat \textit{et al}. \citep{1999PhLB..447..246B,barrow1999cosmologies,avelino1999does,moffat2001acceleration,dkabrowski2013regularizing} to figure out some cosmological problems. Beyond the above problems, the effect of the VSL theory also can be evidenced in some fundamental physics. For example, in 1999, \citet{webb1999search} observed the variation of fine-structure constant $\alpha=e^2/\hbar c$ with cosmic time from the quasar absorption spectra. In following works, \citet{webb2001further} and \citet{king2012spatial} even showed that the fine-structure constant may vary along the space.  We note that the fine-structure constant $\alpha$ is inversely proportional to the speed of light. Therefore, a variation of speed of light also can explain the change of fine-structure constant \citep{barrow1999cosmologies,avelino1999does}. The VSL theory, no doubt, should deserve more attention.

However, there are also some crucial criticisms against the VSL theory. \citet{ellis2007note} argued that VSL theory makes much of modern physics which depends on a constant speed of light to be rewritten, and also pointed out a number of detailed issues due to the VSL theory, such as the VSL theory might contradict Lorentz invariance, and must modify Maxwell's equation. As the research on this theory develops, some problems have been solved.
One of those problems is the breakage of the Lorentz invariance \citep{bassett2000geometrodynamics}, and Ref. \citep{jafari2004operational} showed that VSL theories in the Fock-Lorentz \citep{fok1959theory,manida1999fock} and Magueijo-Smolin \citep{magueijo2002lorentz,magueijo2003generalized} transformations are the redescriptions of special relativity. Thus, the Lorentz invariance is not violated. However, the VSL theory also has to face some other serious problems. In addition, there is not any evidence to support, or completely eliminate this theory. The VSL cosmology is still not universally accepted.

In summary, the VSL theory can successfully solve some significant problems, but it is still excluded by mainstream physics because it has to face a series of problems. Thus, whether the speed of light is varying cannot be judged by theoretical analysis and should be judged only by investigations of the observational data. This topic is exactly what we want to do. In this paper, we examine VSL theory from the observational data, and from the fundamental physics. The observational data we use here are common Type Ia supernovae (SNIa), cosmic microwave background (CMB), observational $H(z)$ data (OHD), and baryon acoustic oscillations (BAO). In previous works, some of them mainly replaced the dark energy with the effect of variation of the speed of light to explain the cosmic acceleration \citep{2014MPLA...2950103Z}. Now, we should make it clear whether the speed of light is varying from the observational data. Following previous investigations, we select the common form $c(t)=c_0a^n(t)$ with the contribution of dark energy and matter, and consequently construct a varying speed of light dark energy model (VSLDE).  In addition, we also use the $Om(z)$ diagnostic to discriminate the varying speed of light dark energy model from the $\Lambda$ cold dark matter ($\Lambda$CDM). The method can be regarded as a criterion to test whether the speed of light is varying.

This paper is organized as follows. In Sec. \ref{VSL}, the VSL theory and the specific form of the speed of light are presented. The method for constraining cosmological models with SNIa, CMB, OHD, and BAO is introduced in Sec. \ref{data}. In Sec. \ref{constrain}, we present the constraint results from the observational data. In Sec. \ref{om}, the $Om(z)$ diagnostic is presented and applied to discriminate the VSL model from $\Lambda$CDM. Finally, the summary and discussion are given in Sec. \ref{conclusion}.

\section{The varying speed of light theory}
\label{VSL}

First of all, we can introduce the time variable $c$, i.e., $c \rightarrow c(t)$. Generally, one can introduce the time variable $c(t)$ in the metric or in the Friedmann equations. In Ref. \citep{bassett2000geometrodynamics}, they believe that the introduction of $c(t)$ in the metric does not change the physics by a coordinate transformation $c \, dt_{\textmd{new}} = c(t) \, dt$. Actually, the scalar factor $a(t)$ will be $a(t_{\textmd{new}})$, which will bring a series of confusing results.  In addition, as noted in Ref. \citep{ellis2007note}, the speed of light $c$ is not a dimensionless quantity. Consequently, one can change the quantity $c$ to any value wanted with the transformations of coordinates. The new speed of light $c$ in the new coordinate is just the coordinate speed of light. In conclusion, in order not to change the reality of cosmic time, we would like to investigate the behavior of VSL theory in the normal cosmic time frame. Thus, the introduction of the time variable $c(t)$ in the metric is more fundamental than that in the Friedmann equations, which is used by many works about VSL theory. Consequently, more fundamental characteristics of the VSL cosmology will be revealed, which will be discussed in detail in the following.

At first, we adopt the cosmological principle that implies a homogenous and isotropic universe, and assume the speed of light is a function of cosmic time. The Friedmann-Robertson-Walker(FRW) metric can be written as
\begin{equation} \label{metric}
ds^{2}=-c^2(t)dt^{2}+a^{2}\left[\frac{dr^{2}}{1-kr^2}+r^{2}\left(d\theta^{2}+sin^{2}\theta d\phi^{2}\right)\right],
\end{equation}
where $k$ denotes the curvature of the space. We will just think about a spatial flat FRW universe, namely $k=0$, for simplicity. The Einstein field equation reads
\begin{equation}\label{Ein}
G_{\mu\nu}=\frac{8\pi G}{c^4(t)}T_{\mu\nu}.
\end{equation}
By substituting Eq. (\ref{metric}) into Eq. (\ref{Ein}), we can obtain the Friedmann equations with a varying speed of light,
\begin{eqnarray}\label{Fried}
3\frac{\dot{a}^2}{a^2} & = & 8\pi G\rho,\\
2\frac{\ddot{a}}{a}+\frac{\dot{a}^2}{a^2}-2\frac{\dot{a}}{a}\frac{\dot{c}}{c} & = & -\frac{8\pi G}{c^2}p, \label{Fried2}
\end{eqnarray}
where $\rho$ and $p$ are the total energy density and pressure, respectively. Here the dot denotes the derivative with respect to cosmic time $t$. Obviously, the term $-2\frac{\dot{a}}{a}\frac{\dot{c}}{c}$ is the effect of the varying speed of light compared with common Friedmann equations. Meanwhile, this term is the difference between the introduction of $c(t)$ in the metric and in Friedmann equations. As we mentioned above, such Friedmann Eqs. (\ref{Fried}) and (\ref{Fried2}), should be more fundamental. The significant effect of this term will be revealed in the following calculations. The corresponding conservation equation reads
\begin{equation}  \label{conservation}
\dot{\rho}+3H\left(\rho+\frac{p}{c^2}\right)=\frac{3H^2}{4\pi G}\frac{\dot{c}}{c},
\end{equation}
where $H=\dot{a}/a$ is the Hubble parameter. Equation (\ref{conservation}) shows that the energy-momentum conservation is broken for $\dot{c}(t) \neq 0$. And any change in the speed of light is a source of matter creation. In fact, we can also obtain the same conclusion from Eq. (\ref{Ein}) as
\begin{equation}
\left(\frac{8\pi G}{c^4(t)}T^{\mu\nu}\right)_{;\mu}\neq 0.
\end{equation}
To solve this problem, the following two solutions have been proposed. First, it is to modify the right side of Eq.(\ref{Ein}). We can add other terms to $T^{\mu\nu}$ \citep{shojaie2006cosmology} or make gravitational constant $G$ vary in time, so that $G(t)c(t)^{-4}=$const \citep{barrow1999cosmologies}. Thus, the energy-momentum conservation can be satisfied. Second, we can neglect the energy-momentum conservation, and regard the variation of the speed of light as a source of matter creation. More related research has been discussed in Refs. \citep{shojaie2006cosmology,shojaie2007varying,clayton1999dynamical}. In this paper, we adopt the latter suggestion.

The following form for the speed of light has been widely used \citep{barrow1999cosmologies}:
\begin{equation}\label{ct}
c(t)=c_0a^n(t)=c_0\left(1+z\right)^{-n},
\end{equation}
where $c_0$ is the current value of the speed of light, and $n$ is a constant. It is obvious that the speed of light $c(t)\rightarrow c_0$ with $n\rightarrow 0$. Moreover, we have $\dot{c}/c=n\dot{a}/a$, so the speed of light grows from zero to $c_0$ for $n>0$, and decreases from infinity to $c_0$ for $n<0$. As discussed in Ref. \citep{2014arXiv1406.0150B}, \citeauthor{2014arXiv1406.0150B} investigated the influence of VSL on the luminosity distance. From the cosmographic method, they found that negative $n$ would lead the universe to accelerate stronger. Positive $n$ would lead to the stronger deceleration. Moreover, a precise test of the fine-structure constant \citep{2007MNRAS.378..221M} provides that the constant $n$ is rather small. Therefore, the parameter $n$ is a considerable significant factor that describes the varying of the speed of light or the evolution of the universe.

Using Eqs. (\ref{Fried}) and  (\ref{ct}), the conservation equation (\ref{conservation}) could be rewritten as
\begin{equation}
\dot{\rho}+3H\left(\rho+\frac{p}{c^2}\right) = 2n \rho H.
\end{equation}
We can define the equation of state (EoS) as
\begin{equation}\label{es}
w \equiv \frac{p}{\rho c^2(t)}=w_0 a^{-2n}=w_0 \left(1+z\right)^{2n},
\end{equation}
where $w_0\equiv p/\rho c_0^2$ denotes the current value of EoS. For matter $w_{m0}=0$, and for radiation $w_{r0}=1/3$. And then, the conservation equation is
\begin{equation}\label{cons}
\dot{\rho}+\rho H\left(3+3w -2n\right)=0.
\end{equation}
Solving this differential equation, we have
\begin{equation}\label{rho}
\rho=\rho_0\left(1+z\right)^{-2n}\exp\left[\int^{z}_{0}\frac{3(1+w)}{1+z}dz\right],
\end{equation}
where $\rho_0$ is the present value of the total energy density. If we only consider the contribution of dark energy and matter, namely,  $\rho=\rho_m+\rho_x$, where $\rho_m$ denotes matter density and $\rho_x$ is dark energy density, Eq. (\ref{rho}) can be expanded as
\begin{eqnarray}
\rho_m &=& \rho_{m0}\left(1+z\right)^{3-2n}, \\
\rho_x &=& \rho_{x0}\left(1+z\right)^{-2n}\exp\left[\int^{z}_{0}\frac{3(1+w_x)}{1+z}dz\right],
\end{eqnarray}
where $w_x=w_{x0}\left(1+z\right)^{2n}$, and $w_{x0}$ is the current value of the equation of state of dark energy. When $n \rightarrow 0$, all of these equations will reduce to the general dark energy model. Solving the corresponding Friedmann equation (\ref{Fried}), we obtain the Hubble parameter
\begin{equation}\label{Ez}
\begin{aligned}
E^2=&\Omega_{m0}\left(1+z\right)^{3-2n}\\
&+\Omega_{x0}\left(1+z\right)^{-2n}\exp\left[\int^{z}_{0}\frac{3(1+w_x)}{1+z}dz\right],
\end{aligned}
\end{equation}
where $E^2\equiv H^2/H_0^2$, $\Omega_{m0}$ and $\Omega_{x0}$ are the matter and dark energy density parameter today, respectively. From the normalization condition, we have three independent variables, $\Omega_{m0}$, $w_{x0}$, and $n$, which should be confirmed by the observational data.

\section{The method for constraining cosmological models and observational data}
\label{data}

In this section, we would like to introduce the observational data and constraint method. The corresponding observational data we use are distance moduli of SNIa, CMB shift parameter, OHD Hubble parameter, and BAO distance parameter.  However, we should notice that the varying speed of light would affect the distance measurements. As found by \citet{barrow2000can}, the variable $c$ does not introduce an intrinsic effect in the spectral line of supernova and also does not dim the standard candle. However, the variable $c$ would correct the Hubble diagram, which induces that the objects in the VSL theory are farther away from us because of the faster speed of light. This effect also can be evidenced in the Taylor series expansion of luminosity distance \citep{barrow2000can,balcerzak2013redshift}. As shown by them, this effect is presented by the introduction of $n$ in the second term. In addition, the variation of the speed of light would affect the evolution of cosmology through the varying EoS of dark energy.

\subsection{Type Ia supernovae}  \label{SNIa data}

As early as 1998, cosmic
accelerating expansion was first observed by  ``standard candle"
SNIa which all have the same intrinsic luminosity. Therefore, the
observables are usually presented in the distance modulus, the
difference between the apparent magnitude $m$ and the absolute
magnitude $M$. The latest version is the Union2.1 compilation
\citep{suzuki2012hubble} which includes 580 samples. They are
discovered by the Hubble Space Telescope Cluster Supernova Survey
over the redshift interval $0.01<z<1.42$. The theoretical
distance modulus is given by
\begin{equation}
\mu_{th}(z)= m-M= 5 \log_{10}D_L(z)+\mu_0,
\end{equation}
where $\mu_0=42.38-5\log_{10}h$, and $h$ is the Hubble
constant $H_0$ in units of 100 km s$^{-1}$Mpc$^{-1}$. The
corresponding luminosity distance function $D_L(z)$  in the VSL theory is
\begin{eqnarray}\label{dl}
 D_L(z) = (1 + z) \int_{0}^{z} \frac{ c(t)  dz'}{E(z'; \textbf{p})},
\end{eqnarray}
where $E(z'; \textbf{p})$ is the dimensionless Hubble parameter given by Eq. (\ref{Ez}), and \textbf{p} stands for the parameter vector of the evaluated model embedded in expansion rate parameter $E(z)$. We note that the parameters in the expansion rate $E(z)$ include the annoying parameter $h$. To be immune from the Hubble constant, we should marginalize over the nuisance parameter $\mu_0$  by integrating the probabilities on $\mu_0$ \citep{pietro2003future,nesseris2005comparison,perivolaropoulos2005constraints}. Finally, we can estimate
the remaining parameters without $h$ by minimizing
\begin{equation} \label{chi2_SN2}
    \tilde{\chi}^{2}_{\rm SN}(z,\textbf{p})= A - \frac{B^2}{C},
\end{equation}
where
\begin{eqnarray}
     A(\textbf{p}) &=&  \sum_{i} \frac{[\mu_{obs}(z) - \mu_{th}(z; \mu_0 = 0,
     \textbf{p})]^2}{\sigma_{i}^{2}(z)},   \nonumber\\
     B(\textbf{p}) &=&  \sum_{i} \frac{\mu_{obs}(z) - \mu_{th}(z; \mu_0=0,
     \textbf{p})}{\sigma_{i}^{2}(z)},   \nonumber\\
     C &=&  \sum_{i} \frac{1}{\sigma_{i}^{2}(z)} , \nonumber
\end{eqnarray}
and $\mu_{obs}$ is the observational distance modulus. This approach has been used in the reconstruction of dark energy
\citep{wei2007reconstruction}, parameter constraint
\citep{wei2010observational},  reconstruction of the energy
condition history \citep{wu2012reconstructing}, etc.

\subsection{Cosmic microwave background}  \label{cmb data}

The CMB experiment measures the temperature and
polarization anisotropy of the cosmic radiation in the early
epoch. It generally plays a major role in establishing and
sharpening the cosmological models. In the CMB measurement, the shift parameter
$R$ is a convenient way to quickly evaluate the likelihood
of the cosmological models. It can be obtained from acoustic oscillations \citep{hu1996small,hinshaw2009five} and contains the main information of the CMB observation. In the VSL theory, it is expressed as
\begin{equation}
R=\sqrt{\Omega_{m0}}\int^{z_s}_0\frac{ c(t)  dz'}{E(z';\textbf{p})},
\end{equation}
where $z_s=1090.97$ is the redshift of decoupling.  According to the measurement of
WMAP-9 \citep{hinshaw2012nine}, we estimate the parameters by minimizing the corresponding
$\chi^2$ statistics
    \begin{equation}  \label{cmb constraint}
     \chi ^2_R = \left( \frac{R-1.728}{0.016} \right)^2.
     \end{equation}

\subsection{Observational $H(z)$ data} \label{ohd}

Unlike the distance measurement, the OHD is an indirect measurement of the cosmic expansion history. The $H(z)$ data can be obtained from differential ages of galaxies \citep{simon2005constraints,stern2010cosmic}, from the BAO peaks in the galaxy power spectrum
\cite{gaztanaga2009clustering,moresco2012improved} or from the BAO
peaks using the Ly$\alpha$ forest of quasars \cite{2013A&A...552A..96B}. In this paper, we will use the new data from
Ref. \citep{farooq2013hubble} that contains 28 observational data. The best-fit values of the parameters can be obtained by minimizing
\begin{equation}
\chi^2_{\rm{OHD}}=\sum_{i}^{28}\frac{[H_0E(z_i)-H_{obs}(z_i)]^2}{\sigma_i^2}.
\end{equation}
In the calculation, we adopt the WMAP9 result \citep{hinshaw2012nine}  $H_0=70\pm 2.2$ km s$^{-1}$Mpc$^{-1}$ as the prior.

\subsection{Baryon acoustic oscillation}   \label{bao data}

The measurement of BAO in the large-scale galaxies has rapidly become one of the most important observational pillars in cosmological constraints. This measurement is usually called the standard ruler in cosmology \citep{eisenstein1998baryonic}.
The distance parameter $A$ obtained from the BAO peak in the distribution of sloan digital sky survey luminous red galaxies \citep{eisenstein2005detection} is a significant parameter and is defined as
\begin{equation}
A_{th}=\Omega_{m0}^{1/2}E(z_1)^{-1/3}\left[\frac{1}{z_1}\int^{z_1}_{0}\frac{c(t)  dz'}{E(z';\textbf{p})}\right]^{2/3}.
\end{equation}
We use the three combined data points in Ref. \citep{addison2013cosmological} that cover $0.1 < z <2.4$ to determine the parameters in evaluated models. The expression of $\chi^2$ statistics is
\begin{equation}
\chi^2_{A}=\sum_{i} \left(\frac{A_{th}-A_{obs}}{\sigma^2_A}\right)^2,
\end{equation}
where $A_{obs}$ is every observational distance parameter and $\sigma_A$ is its corresponding error.

Since the SNIa, CMB, OHD, and BAO data points are effectively
independent measurements, we can simply minimize their total
$\chi^{2}$ values as
\begin{eqnarray}
\label{21}\chi^2(z, \textbf{p})= \tilde{\chi}^{2}_{\rm SN} + \chi^{2}_{R}+\chi^{2}_{
A}+\chi^2_{\rm{OHD}}, \nonumber
\end{eqnarray}
to determine the parameters in the VSLDE.

\section{constraining the parameters of VSL theory with observational data}
\label{constrain}
\begin{figure}
\begin{center}
\includegraphics[width=0.48\textwidth]{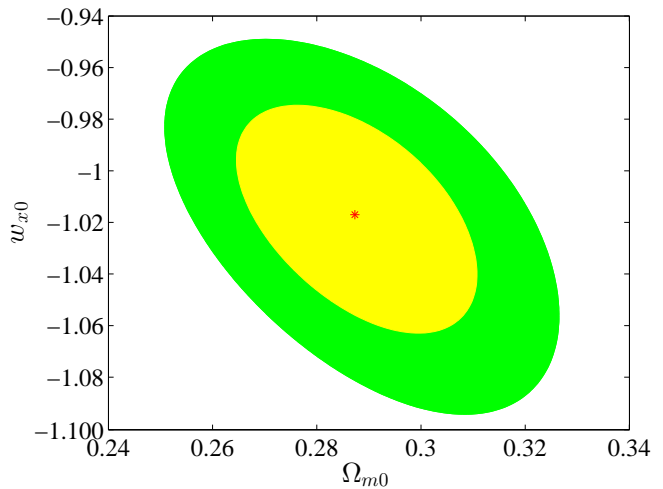}
\caption{The $68.3\%$ and $95.4\%$ confidence regions in the $w_{x0}$-$\Omega_{m0}$ parameter space for VSLDE, which are constrained by combined observational data of SNIa, CMB, OHD and BAO. The red asterisk indicates the best-fit point. \label{omwc}}
\end{center}
\end{figure}

\begin{figure}
\begin{center}
\includegraphics[width=0.48\textwidth]{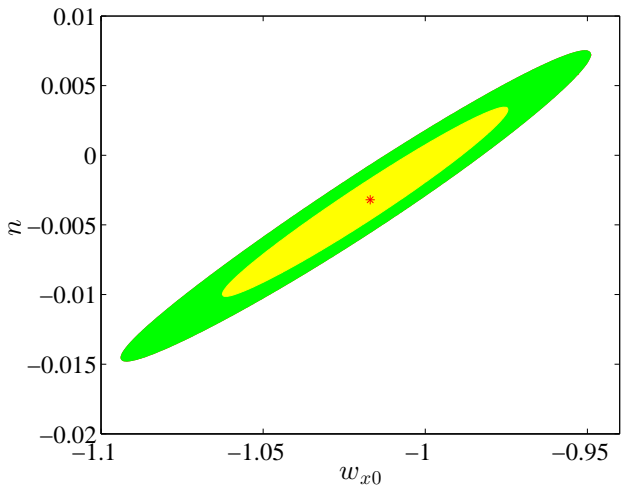}
\caption{Same as Fig. \ref{omwc} but for the $n$-$w_{x0}$ parameter space. \label{wcn}}
\end{center}
\end{figure}

\begin{figure}
\begin{center}
\includegraphics[width=0.48\textwidth]{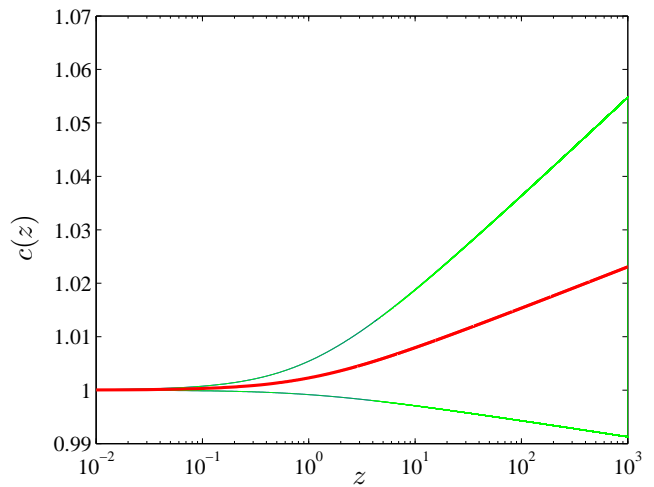}
\caption{The reconstruction of speed of light $c (z)$. The central red solid line is the best-fit value of $c(z)$. Top and bottom lines are the errors of reconstructed $c(z)$ in the $1\sigma$ confidence level. \label{speed}}
\end{center}
\end{figure}

\begin{figure}
\begin{center}
\includegraphics[width=0.48\textwidth]{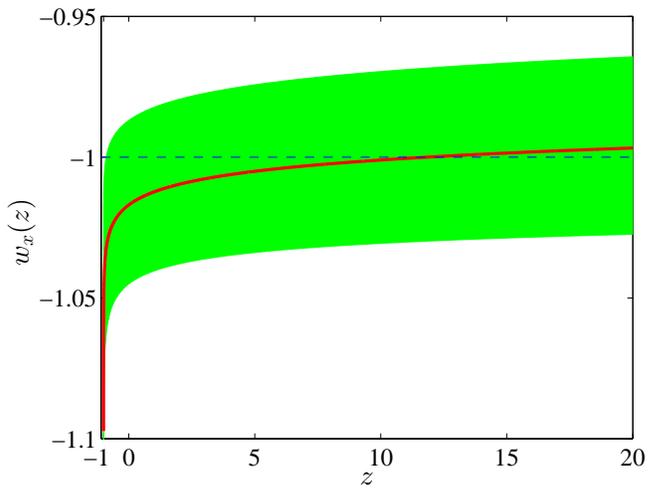}
\caption{The reconstruction of EoS of dark energy $w_x (z)$. The central red solid line is the best-fit value of $w_x (z)$. The shaded regions are the errors of reconstructed EoS within $1\sigma$ confidence level. \label{wz}}
\end{center}
\end{figure}

Using the combined observational data sets of SNIa, CMB, OHD, and BAO, we perform the $\chi^2$ statistics. The contour diagram of parameters $w_{x0}$-$\Omega_{m0}$  and   $n$-$w_{x0}$ are given in Figs. \ref{omwc} and \ref{wcn}, respectively.
It is shown that the combined data can present very compact constraints on the parameters.
Considering the degrees of freedom (dof), the result is also pretty good, i.e., $\chi^2_{min}$/dof=0.9602 with parameters $w_{x0}$-$\Omega_{m0}$, and $\chi^2_{min}$/dof=0.9517 with parameters $n$-$w_{x0}$. Marginalizing over the posterior probability, we obtain the matter density $\Omega_{m0}=0.2878^{+0.0153}_{-0.0151}\left(1\sigma\right)$, the current EoS of dark energy $w_{x0}=-1.017^{+0.028}_{-0.030}\left(1\sigma\right)$, and the key constant $n=-0.0033^{+0.0045}_{-0.0045}\left(1\sigma\right)$. As previously mentioned, $n$ is characteristic for the
severity of the varying speed of light. First, from the constrained constant $n$, we find that the variation of the speed of light is tiny. The constant $n$ is only in order of 10$^{-3}$. Moreover, the observational data show that nonvariable $c$ cannot be ruled out in the $1\sigma$ confidence level. In Fig. \ref{speed}, we reconstruct $c(z)$ with redshift $z$. We find that the speed of light decreases with decreasing redshift $z$. Considering the error estimation of $c(z)$, we obtain that the speed of light almost has no variation for $z<10^{-1}$. Even in the early epoch, its variation is only in order of 10$^{-2}$. In addition, we also deduce that the high-$z$ data are more sensitive to the VSL model than the lower-$z$ data. Although current data including the CMB cannot remove or admit the VSL model with a high significance, the future observations with a high redshift may be able to test this effect with the improvement of observational precision. Taking the CMB shift parameter as an example, the uncertainty is now 0.9\% from WMAP-9. Besides the parameter $n$, we note that the constrained EoS of dark energy is also interesting. Although the cosmological constant as a candidate of dark energy cannot be ruled out by the data, we find that they favor the phantomlike dark energy with $w_x<-1$ much more. In Fig. \ref{wz}, we show the reconstruction of $w_x (z)$ with the redshift. We find that the best-fit value of EoS is less than $-1$ for $z<10$. For the future, it almost behaves like the phantom. Moreover, the values of the parameters $\Omega_{m0}$ and $w_{x0}$ are pretty consistent with the mainstream physics.

\section{Difference between VSLDE and $\Lambda$CDM}
\label{om}

\begin{figure}
\begin{center}
\includegraphics[width=0.48\textwidth]{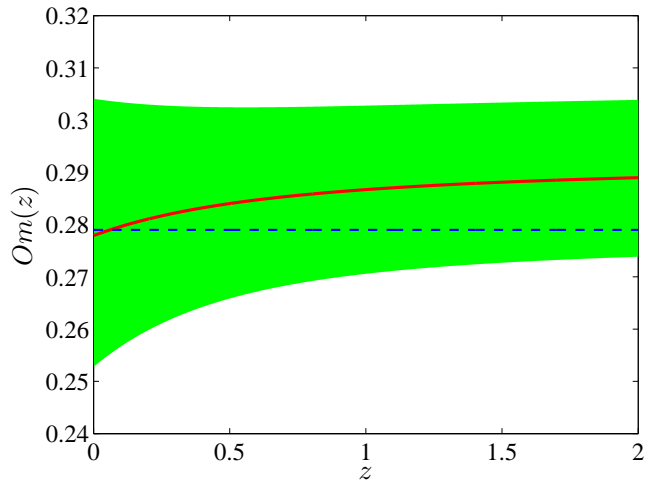}
\caption{The reconstruction of $Om(z)$ diagnostic. The central red solid line is the best-fit curve of $Om(z)$ in the VSLDE. The blue dashed line represents the $Om(z)$ in the spatially flat $\Lambda$CDM. The shaded area shows the 1$\sigma$ error region in the VSLDE. \label{wucha}}
\end{center}
\end{figure}
According to the above analysis, we can see that the VSLDE is very close to $\Lambda$CDM. Now, we apply an effective geometrical diagnostic, $Om(z)$, to directly show the difference between the VSLDE and $\Lambda$CDM. $Om(z)$ is defined as \citep{sahni2008two}
\begin{equation}
Om(z)\equiv \frac{E^2-1}{(1+z)^3-1}.
\end{equation}
For a spatially flat radiation vanished $\Lambda$CDM model, we obtain $Om(z)=\Omega_{m0}$, a constant that is independent of the redshift. Therefore, it is used widely to discriminate some exotic dark energy models (e.g., quintessence \citep{sahni2008two}, phantom \citep{sahni2008two}, and Ricci dark energy model \citep{feng2008statefinder}) from $\Lambda$CDM. Investigations in Ref. \citep{sahni2008two} also show that a positive slope of $Om(z)$ indicates a phase of phantom ($w<-1$) while a negative slope represents quintessence ($w>-1$).

Using the observational data of SNIa, CMB, OHD, and BAO, we obtain $\Omega_{m0}=0.279$ in the $\Lambda$CDM model. Therefore, $Om (z)$ in the $\Lambda$CDM model is the constant 0.279. In Fig. \ref{wucha}, we reconstruct the geometrical diagnostic $Om(z)$ at $68.3\%$ confidence level, and find that the VSLDE and $\Lambda$CDM cannot be discriminated. Because the key factor $n$ in the varying speed of light theory is so feeble. To some extent, the speed of light may be a constant with high significance.

\section{Conclusion and discussion}
\label{conclusion}

Varying speed of light theory has been put forward to explain some cosmological problems, such as horizon, flatness and cosmological singularity problems. However, the VSL cosmology is excluded by mainstream physics because it makes much of modern physics depending on a constant speed of light to be rewritten. Moreover, it also faces a number of other detailed issues. In the present paper, we devote ourselves to testing whether the speed of light is varying from the observational data. Considering a specific form for the variable $c(t)=c_0a^n(t)$, we deduce the corresponding  Friedmann equation and Hubble parameter. In some previous references, some of them introduced the time variable $c(t)$ just in the Friedmann equation. However, we think that the application of $c(t)$ in the metric is much more fundamental and realistic. Using the combined  observational data of SNIa, CMB, OHD, and BAO, we obtain  $\Omega_{m0}=0.2878^{+0.0153}_{-0.0151}$, $w_{x0}=-1.017^{+0.028}_{-0.030}$, and $n=-0.0033^{+0.0045}_{-0.0045}$ at the $1\sigma$ confidence level. From the reconstructed $c(z)$ in Fig. \ref{speed}, we obtain that the speed of light almost has no variation for $z<10^{-1}$. Even in the early epoch, its variation is only in the order of $10^{-2}$. In Ref. \citep{2003PhRvD..68f3511S}, \citeauthor{2003PhRvD..68f3511S} constructed the phase-space portraits of the VSL models and obtained a global view on their dynamics. The most probability is that the universe can be created with constant $c (n=0)$, if the classical spacetime emerges via the quantum tunneling process. Moreover, the least size $a_{0}$ of the VSL FRW universe probably permits an insignificant $-1<n<0$. The speed of light $c$ decreases with the evolution of the universe. Obviously, our investigation verifies and favors the previous deductions.
In addition, we also find that the high-$z$ data are more sensitive to the VSL model than the lower-$z$ data. Applying an effective geometrical diagnostic, $Om(z)$, we find that it is still difficult to differentiate the VSLDE from the $\Lambda$CDM model, because the key factor $n$ is so trivial. Apparently the observational data favor the constant speed of light.

However, we should note that high-$z$ observations are more useful for testing this model. With the improvement of observational precision and the introduction of some new observations, the constrained results may be consolidated. For example, the redshift drift proposed by \citet{sandage1962change} would monitor the wavelength shift of a quasar from the Ly$\alpha$ absorption lines \citep{loeb1998direct,liske2008cosmic}. The measurements can extend to the redshift region $z=2-5$ to get the signal in order of about $10^{-11}$.  In Ref. \citep{balcerzak2013redshift}, the authors investigate the redshift drift in the VSL theory and find that the redshift drift for 15 years can test $|n|<0.045$. Therefore, we can obtain that more observational baselines are needed to distinguish the VSL models from constant-$c$ cosmology. Besides the secular redshift drift, we note that the future gravitational wave probes will have higher sensitivity. With the detection of the B-mode from CMB by the BICEP2 group \cite{ade2014bicep2}, the gravitational wave probes will open a new window for us. The related research can be found in Ref. \citep{2014arXiv1404.5567M}.
In addition, we find from Fig. \ref{wcn} that parameters space $n$ and $w_{x0}$ are related in the considered observational data. Therefore, we also can expect that the inclusion of some other data, such as the gravitational lens and the growth rate, could give a tighter constraint on the VSL model because of different degeneracies.

On the other hand, there exist some deficiencies to advance. In the present paper, we adopt a widely used form of VSL, i.e., $c(t) = c_0 a^{n} (t)$. The conclusion is under this assumption. Therefore, we must be careful to extrapolate our conclusions. In the future, we would like to investigate the VSL theory in other forms.

\acknowledgments{
We quite appreciate the anonymous referee for his/her suggestions to improve this manuscript.
J.-Z. Qi would like to thank Rong-Jia Yang for his useful help. This work is supported by the National Natural Science Foundation of China (Grant No. 11235003, No. 11175019, and No. 11178007).}

\bibliography{citep}
\end{document}